\documentclass[groupedaddress, superscriptaddress, citesort]{revtex4}
\usepackage{amssymb, amsmath}
\usepackage[dvips]{graphicx}

\def\be{\begin{equation}}
\def\ee{\end{equation}}
\def\ba{\begin{array}}
\def\ea{\end{array}}

\newcommand{\RNum}[1]{\uppercase\expandafter{\romannumeral #1\relax}}

\begin{document}
\baselineskip=18pt

\title {Quantum coherence in mutually unbiased bases}
\author{Yao-Kun Wang}
\affiliation{College of Mathematics,  Tonghua Normal University, Tonghua, Jilin 134001, China}
\affiliation{Research Center for Mathematics, College of Mathematics, Tonghua Normal University, Tonghua, Jilin 134001, China}
\author{Li-Zhu Ge}
\affiliation{The Branch Campus of Tonghua Normal University, Tonghua, Jilin 134001, China}
\author{Yuan-Hong Tao}
\affiliation{Department of Mathematics College of Sciences, Yanbian University, Yanji 133002, China}

\begin{abstract}
We investigate the $l_{1}$ norm of coherence of quantum states in mutually unbiased bases. We find that the sum of squared $l_{1}$ norm of coherence of the mixed state single qubit is less than two. We derive the $l_{1}$ norm of coherence of three classes of $X$ states in nontrivial mutually unbiased bases for $4$-dimensional Hilbert space is equal. We proposed ``autotensor of mutually unbiased basis(AMUB)" by the tensor of mutually unbiased bases, and depict the level surface of constant the sum of the $l_{1}$ norm of coherence of Bell-diagonal states in AMUB. We find the $l_{1}$ norm of coherence of Werner states and isotropic states in AMUB is equal respectively.
\end{abstract}

\maketitle

\section{Introduction}

Quantum coherence is a special feature of quantum mechanic like entanglement and other quantum correlations. Quantum coherence is an essential factor in quantum information processing \cite{Jha,Bagan,Kammerlander}, quantum optics \cite{Glauber,Sudarshan,Mandel}, quantum metrology \cite{Giovannetti,Demkowicz,Giovannetti1}, low-temperature thermodynamics \cite{Lostaglio,Lostaglio1,Vazquez,Wacker,pati,aberg,Narasimhachar,Oppenheim}  and quantum biology \cite{Lloyd,Li,Huelga,levi,Plenio,Rebentrost}.
Recently, a structure to quantify coherence has been proposed \cite{Baumgratz}, and various quantum coherence measures, such as the $l_{1}$ norm of coherence \cite{Baumgratz}, trace norm of coherence \cite{shao}, relative entropy of coherence \cite{Baumgratz},   Tsallis relative $\alpha$ entropies \cite{Rastegin} and Relative R\'{e}nyi $\alpha$ monotones \cite{Chitambar}, have been defined.
 With the help of of the coherence measures, a variety of properties of quantum coherence, such as the relations between quantum correlations and quantum coherence \cite{Ma,Streltsov,Radhakrishnan,Yao,Xi}, the freezing phenomenon of coherence \cite{Bromley,Yu}, have been studied.

Mutually unbiased bases are used in detection of quantum
entanglement \cite{Spengler}, quantum state reconstruction \cite{Wootters}, quantum error correction \cite{Gottesman,Calderbank}, and the mean king¡¯s problem \cite{Vaidman,Englert}. Many features of mutually unbiased bases are reviewed in
reference \cite{Durt}. When d is power of a
prime number, maximal sets of $d+1$ mutually unbiased bases have been built for the case. Maximal sets of MUBs are an open problem \cite{Durt}, when the dimensionality is another composite number.
Entropic uncertainty relations for $d+1$ mutually unbiased bases in d-dimensional Hilbert space were obtained in
references \cite{Ivanovic,sanchez}. The fine-grained uncertainty relation for mutually unbiased bases is derived in \cite{ren}. The relation between mutually unbiased bases and unextendible maximally entangled is investigated in \cite{chen}.

In this article, we investigate the  $l_{1}$ norm of coherence of quantum states in mutually unbiased bases. We evaluate analytically the sum of squared $l_{1}$ norm of coherence of the mixed state single qubit. We derive the relation of the $l_{1}$ norm of coherence of three classes of $X$ states in nontrivial mutually unbiased bases for $4$-dimensional Hilbert space. We propose ``autotensor of mutually unbiased basis(AMUB)" by the tensor of mutually unbiased bases, and depict the level surface \cite{lang} of constant the sum of the $l_{1}$ norm of coherence of Bell-diagonal states in AMUB. We obtain the relations of the $l_{1}$ norm of coherence of Werner states and isotropic states in AMUB respectively.

\section{The $l_{1}$ norm of coherence of quantum states in $2$ dimension mutually unbiased bases}

Under fixed reference basis, the $l_{1}$ norm of coherence of state $\rho$ is defined by
\begin{eqnarray}\label{l1}
C_{l_{1}}(\rho)=\sum_{i\neq j}|\rho_{i,j}|,
\end{eqnarray}
and the relative entropy of coherence is given by
\begin{eqnarray}\label{rel}
C_{r}(\rho)=S(\rho_{diag})-S(\rho),
\end{eqnarray}
where $S(\rho)=-Tr{\rho\log\rho}$ is von Neumann entropy.

A set of orthonormal bases $\{B_{k}\}$ for a Hilbert space
$H=C^{d}$ where $\{B_{k}\}=\{|0_{k}\rangle,\cdot \cdot \cdot,|d-1_{k}\rangle$ is
called mutually unbiased (MU) iff
\begin{eqnarray}
|\langle i_{k}|j_{l}\rangle|^{2}=\frac{1}{d}, \forall i,j\in\{0,\cdot \cdot \cdot,d-1\},
\end{eqnarray}
holds for all basis vectors $|i_{k}\rangle$ and $|j_{l}\rangle$ that belong to
different bases, i.e. $\forall k\neq l$.

In dimension $d=2$, a set of three mutually unbiased
bases is readily obtained from the eigenvectors of the
three Pauli matrices $\sigma_{z}$, $\sigma_{x}$ and $\sigma_{y}$:
\begin{eqnarray}
\alpha_{1}&=&\{\alpha_{11},\alpha_{12}\}=\{|0\rangle,|1\rangle\},\nonumber\\
\alpha_{2}&=&\{\alpha_{21},\alpha_{22}\}=\{\frac{1}{\sqrt{2}}(|0\rangle+|1\rangle),\frac{1}{\sqrt{2}}(|0\rangle-|1\rangle)\},\nonumber\\
\alpha_{3}&=&\{\alpha_{31},\alpha_{32}\}=\{\frac{1}{\sqrt{2}}(|0\rangle+i|1\rangle),\frac{1}{\sqrt{2}}(|0\rangle-i|1\rangle)\}\nonumber.
\end{eqnarray}

In dimension $d=3$, there are four mutually unbiased bases as fowllow:
\begin{eqnarray}
\beta_{1}&=&\{\beta_{11},\beta_{12},\beta_{13}\}=\{|0\rangle,|1\rangle,|2\rangle\},\nonumber\\
\beta_{2}&=&\{\beta_{21},\beta_{22},\beta_{23}\}=\{\frac{1}{\sqrt{3}}(|0\rangle+|1\rangle+|2\rangle),\frac{1}{\sqrt{3}}(|0\rangle+\omega|1\rangle+\omega^{2}|2\rangle),\frac{1}{\sqrt{3}}(|0\rangle+\omega^{2}|1\rangle+\omega|2\rangle)\},\nonumber\\
\beta_{3}&=&\{\alpha_{31},\alpha_{32},\alpha_{33}\}=\{\frac{1}{\sqrt{3}}(|0\rangle+|1\rangle+\omega^{2}|2\rangle),\frac{1}{\sqrt{3}}(|0\rangle+\omega^{2}|1\rangle+|2\rangle),\frac{1}{\sqrt{3}}(|0\rangle+\omega|1\rangle+\omega|2\rangle)\},\nonumber\\
\beta_{4}&=&\{\alpha_{41},\alpha_{42},\alpha_{43}\}=\{\frac{1}{\sqrt{3}}(|0\rangle+|1\rangle+\omega|2\rangle),\frac{1}{\sqrt{3}}(|0\rangle+\omega|1\rangle+|2\rangle),\frac{1}{\sqrt{3}}(|0\rangle+\omega^{2}|1\rangle+\omega^{2}|2\rangle)\},\nonumber
\end{eqnarray}
where $\omega=e^{i\frac{2\pi}{3}}$.

An arbitrary density matrix for a mixed state single qubit may be written as
\begin{eqnarray}
\rho_{s}=\frac{I+\overrightarrow{r}\cdot\overrightarrow{\sigma}}{2}\nonumber
\end{eqnarray}
where $\overrightarrow{r}=(x,y,z)$ is a real three-dimensional vector such that $x^{2}+y^{2}+z^{2}\leq 1$, and $\overrightarrow{\sigma}=(\sigma_{x},\sigma_{y},\sigma_{z})$. In particular, $\rho$ is pure if and only if $x^{2}+y^{2}+z^{2}=1$.

Next, we will consider the relation of the $l_{1}$ norm of coherence among $\rho_{s}$ in three mutually unbiased
bases $\alpha_{1}, \alpha_{2}, \alpha_{3}$.

The density matrix of mixed state single qubit $\rho_{s}$ in base $\alpha_{1}=\{\alpha_{11},\alpha_{12}\}=\{|0\rangle,|1\rangle\}$ is
\begin{eqnarray}\label{s1}
\rho_{s}&=&\frac{1}{2}\left(
\begin{array}{cc}
1+z & x-i y \\
x+iy & 1-z \\
\end{array}
\right)\nonumber\\
&=&\frac{1}{2}(1+z)|0\rangle\langle0|+\frac{1}{2}(x-iy)|0\rangle\langle1|+\frac{1}{2}(x+iy)|1\rangle\langle0|+\frac{1}{2}(1-z)|1\rangle\langle1|,
\end{eqnarray}
Using Eq. (\ref{l1}) directly, the $l_{1}$ norm of coherence of state $\rho_{s}$ in base $\alpha_{1}$ is
\begin{eqnarray}
C_{l_{1}}(\rho_{s})_{\alpha_{1}}=|\frac{1}{2}(x-iy)|+|\frac{1}{2}(x+iy)|=\sqrt{x^{2}+y^{2}}.
\end{eqnarray}

The density matrix of  $\rho_{s}$ in base $\alpha_{2}=\{\alpha_{21},\alpha_{22}\}$ is
\begin{eqnarray}\label{s2}
\rho_{s}&=&\left(
\begin{array}{cc}
a_{11} & a_{12} \\
a_{21} & a_{22} \\
\end{array}
\right)\nonumber\\
&=&a_{11}\alpha_{21}\alpha^{\dag}_{21}+a_{12}\alpha_{21}\alpha^{\dag}_{22}+a_{21}\alpha_{22}\alpha^{\dag}_{21}+a_{22}\alpha_{22}\alpha^{\dag}_{22}\nonumber\\
&=&\frac{1}{2}(a_{11}+a_{12}+a_{21}+a_{22})|0\rangle\langle0|+\frac{1}{2}(a_{11}-a_{12}+a_{21}-a_{22})|0\rangle\langle1|\nonumber\\
& &+\frac{1}{2}(a_{11}+a_{12}-a_{21}-a_{22})|1\rangle\langle0|+\frac{1}{2}(a_{11}-a_{12}-a_{21}+a_{22})|1\rangle\langle1|.
\end{eqnarray}
As $\rho_{s}$ in Eq. (\ref{s1}) and Eq. (\ref{s2}) is the same, using the method of undeterminated coefficients, we obtain
\begin{equation}
\begin{cases}\nonumber
\frac{1}{2}(a_{11}+a_{12}+a_{21}+a_{22})=\frac{1}{2}(1+z)\\
\frac{1}{2}(a_{11}-a_{12}+a_{21}-a_{22})=\frac{1}{2}(x-iy)\\
\frac{1}{2}(a_{11}+a_{12}-a_{21}-a_{22})=\frac{1}{2}(x+iy)\\
\frac{1}{2}(a_{11}-a_{12}-a_{21}+a_{22})=\frac{1}{2}(1-z)
\end{cases}.
\end{equation}
The solution of the equation is
\begin{equation}
\begin{cases}
a_{11}=\frac{1+x}{2}\\
a_{12}=\frac{z+iy}{2}\\
a_{21}=\frac{z-iy}{2}\\
a_{22}=\frac{1-x}{2}
\end{cases}.
\end{equation}
The $l_{1}$ norm of coherence of state $\rho_{s}$ in base $\alpha_{2}$ is
\begin{eqnarray}
C_{l_{1}}(\rho_{s})_{\alpha_{2}}=|\frac{1}{2}(z+iy)|+|\frac{1}{2}(z-iy)|=\sqrt{z^{2}+y^{2}}.
\end{eqnarray}

The density matrix of  $\rho_{s}$ in base $\alpha_{3}=\{\alpha_{31},\alpha_{32}\}$ by the above method is
\begin{eqnarray}\nonumber\\
\rho_{s}&=&\frac{1}{2}\left(
\begin{array}{cc}
1+y & z-ix \\
z+ix & 1-y \\
\end{array}
\right)
\end{eqnarray}
The $l_{1}$ norm of coherence of state $\rho_{s}$ in base $\alpha_{3}$ is
\begin{eqnarray}
C_{l_{1}}(\rho_{s})_{\alpha_{3}}=|\frac{1}{2}(z-ix)|+|\frac{1}{2}(z+ix)|=\sqrt{z^{2}+x^{2}}.
\end{eqnarray}

As $x^{2}+y^{2}+z^{2}\leq 1$, $[C_{l_{1}}(\rho_{s})_{\alpha_{1}}]^{2}+[C_{l_{1}}(\rho_{s})_{\alpha_{2}}]^{2}+[C_{l_{1}}(\rho_{s})_{\alpha_{3}}]^{2}\leq 2.$

\section{The $l_{1}$ norm of coherence of $X$ states in the tensor of $3$ dimension mutually unbiased bases}

For the three classes of $X$ states in base $\beta_{1}=\{\beta_{11},\beta_{12},\beta_{13}\}=\{|0\rangle,|1\rangle,|2\rangle\}$
\begin{eqnarray}\label{x1}
 \rho_{X}=\left(
            \begin{array}{ccc}
              x & 0 & z \\
              0 & 1-x-y & 0 \\
              z & o & y \\
            \end{array}
          \right),
\end{eqnarray}
where $x, y, z$ are all real number, we will consider the $l_{1}$ norm of coherence of $\rho_{X}$ in the $3$ dimension mutually unbiased bases $\beta_{2}, \beta_{3}, \beta_{4}$.

Let the density matrix of $\rho_{X}$ in base $\beta_{2}=\{\beta_{21},\beta_{22},\beta_{23}\}$ be
\begin{eqnarray}\label{x2}
 \rho_{X}=\left(
            \begin{array}{ccc}
              b_{11} & b_{12} & b_{13} \\
              b_{21} & b_{22} & b_{23} \\
              b_{31} & b_{32} & b_{33} \\
            \end{array}
          \right),
\end{eqnarray}
and $\rho_{X}=b_{11}\beta_{21}\beta_{21}^{\dag}+b_{12}\beta_{21}\beta_{22}^{\dag}+b_{13}\beta_{21}\beta_{23}^{\dag}+b_{21}\beta_{22}\beta_{21}^{\dag}
+b_{22}\beta_{22}\beta_{22}^{\dag}+b_{23}\beta_{22}\beta_{23}^{\dag}+b_{31}\beta_{23}\beta_{21}^{\dag}+b_{32}\beta_{23}\beta_{22}^{\dag}+b_{33}\beta_{23}\beta_{23}^{\dag}$.
As $\rho_{X}$ in Eq. (\ref{x1}) and Eq. (\ref{x2}) is the same, using the method of undeterminated coefficients, we obtain
\begin{equation}
\begin{cases}
\frac{1}{3}(b_{11}+b_{12}+b_{13}+b_{21}+b_{22}+b_{23}+b_{31}+b_{32}+b_{33})=x\\
\frac{1}{3}(b_{11}+\omega^{2}b_{12}+\omega b_{13}+b_{21}+\omega^{2}b_{22}+\omega b_{23}+b_{31}+\omega^{2}b_{32}+\omega b_{33})=0\\
\frac{1}{3}(b_{11}+\omega b_{12}+\omega^{2}b_{13}+b_{21}+\omega b_{22}+\omega^{2}b_{23}+b_{31}+\omega b_{32}+\omega^{2}b_{33})=z\\
\frac{1}{3}(b_{11}+b_{12}+b_{13}+\omega b_{21}+\omega b_{22}+\omega b_{23}+\omega^{2}b_{31}+\omega^{2}b_{32}+\omega^{2}b_{33})=0\\
\frac{1}{3}(b_{11}+\omega^{2}b_{12}+\omega b_{13}+\omega b_{21}+b_{22}+\omega^{2}b_{23}+\omega^{2}b_{31}+\omega b_{32}+b_{33})=1-x-y\\
\frac{1}{3}(b_{11}+\omega b_{12}+\omega^{2}b_{13}+\omega b_{21}+\omega^{2}b_{22}+b_{23}+\omega^{2}b_{31}+b_{32}+\omega b_{33})=0\\
\frac{1}{3}(b_{11}+b_{12}+b_{13}+\omega^{2}b_{21}+\omega^{2}b_{22}+\omega^{2}b_{23}+\omega b_{31}+\omega b_{32}+\omega b_{33})=z\\
\frac{1}{3}(b_{11}+\omega^{2}b_{12}+\omega b_{13}+\omega^{2}b_{21}+\omega b_{22}+b_{23}+\omega b_{31}+b_{32}+\omega^{2}b_{33})=0\\
\frac{1}{3}(b_{11}+\omega b_{12}+\omega^{2}b_{13}+\omega^{2}b_{21}+b_{22}+\omega b_{23}+\omega b_{31}+\omega^{2}b_{32}+b_{33})=y
\end{cases}.
\end{equation}
The solution of the equation is
\begin{equation}
\begin{cases}
b_{11}=\frac{1+2z}{3}, b_{12}=\frac{(3x+z-1)-\sqrt{3}(x+2y+z-1)i}{6}, b_{13}=\frac{(3x+z-1)+\sqrt{3}(x+2y+z-1)i}{6},\\
b_{21}=\overline{b_{12}}, b_{22}=\frac{1-z}{3}, b_{23}=\frac{(3x-2z-1)-\sqrt{3}(x+2y-2z-1)i}{6},\\
b_{31}=\overline{b_{13}}, b_{32}=\overline{b_{23}}, b_{33}=\frac{1-3z}{3}.
\end{cases}
\end{equation}
The $l_{1}$ norm of coherence of state $\rho_{X}$ in base $\beta_{2}$ is
\begin{eqnarray}
C_{l_{1}}(\rho_{X})_{\beta_{2}}=2(|b_{12}|+|b_{13}|+|b_{23}|).
\end{eqnarray}

Similarly, the density matrix of  $\rho_{X}$ in base $\beta_{3}$ is
\begin{eqnarray}
 \rho_{X}=\left(
            \begin{array}{ccc}
              b_{22} & \overline{b_{12}} & b_{23} \\
              b_{12} & b_{11} & b_{13} \\
              \overline{b_{23}} & \overline{b_{13}} & b_{33} \\
            \end{array}
          \right),
\end{eqnarray}
The $l_{1}$ norm of coherence of state $\rho_{X}$ in base $\beta_{3}$ is
\begin{eqnarray}
C_{l_{1}}(\rho_{X})_{\beta_{3}}=2(|b_{12}|+|b_{13}|+|b_{23}|).
\end{eqnarray}

The density matrix of  $\rho_{X}$ in base $\beta_{4}$ is
\begin{eqnarray}
 \rho_{X}=\left(
            \begin{array}{ccc}
              b_{22} & b_{12} & \overline{b_{23}} \\
              \overline{b_{12}} & b_{11} & \overline{b_{13}} \\
              b_{23} & b_{13} & b_{33} \\
            \end{array}
          \right),
\end{eqnarray}
The $l_{1}$ norm of coherence of state $\rho_{X}$ in base $\beta_{4}$ is
\begin{eqnarray}
C_{l_{1}}(\rho_{X})_{\beta_{4}}=2(|b_{12}|+|b_{13}|+|b_{23}|).
\end{eqnarray}

At last, we find that the $l_{1}$ norm of coherence of state $\rho_{X}$ in base $\beta_{2}, \beta_{3}, \beta_{4}$ is equal, i.e
\begin{eqnarray}
C_{l_{1}}(\rho_{X})_{\beta_{2}}=C_{l_{1}}(\rho_{X})_{\beta_{3}}=C_{l_{1}}(\rho_{X})_{\beta_{4}}.
\end{eqnarray}

Furthermore, let
\begin{eqnarray}
 \rho_{\vartriangle}=\left(
            \begin{array}{ccc}
              1-x-y & 0 & 0 \\
              0 & x & z \\
              0 & z & y \\
            \end{array}
          \right),
\end{eqnarray}
and
\begin{eqnarray}
 \rho_{\triangledown}=\left(
            \begin{array}{ccc}
              x & z & 0 \\
              z & y & 0 \\
              0 & 0 & 1-x-y \\
            \end{array}
          \right),
\end{eqnarray}
where $x, y, z$ are all real number, using above method, we can find that the $l_{1}$ norm of coherence of state $\rho_{\vartriangle}$ and $\rho_{\triangledown}$ in base $\beta_{2}, \beta_{3}, \beta_{4}$ is also equal respectively.
\section{The $l_{1}$ norm of coherence of Bell-diagonal states in the tensor of $2$ dimension mutually unbiased bases}

In this section, we extend the concept of mutually unbiased basis by the tensor.

\emph{Definition}. For the set of mutually unbiased bases $\{B_{k}\}$ for a Hilbert space $H=C^{d}$ where $\{B_{k}\}=\{|0_{k}\rangle,\cdot \cdot \cdot,|d-1_{k}\rangle\}$, we call the set $\{\gamma_{k}\}=\{|i\rangle_{k}\otimes|i\rangle_{k}| \forall i, j\in\{0,\cdot\cdot\cdot,d-1\}\}$ \emph{autotensor of mutually unbiased basis}(AMUB) if
\begin{eqnarray}
|(\langle i|_{k}\otimes\langle j|_{k})(|m\rangle_{l}\otimes|n\rangle_{l})|=\frac{1}{d},
\end{eqnarray}
where $k\neq l$.
Furthermore, we can construct a set of AMUB by $d=2$ dimension mutually unbiased bases. For example, let
\begin{eqnarray}
\gamma_{1}&=&\{\gamma_{11},\gamma_{12},\gamma_{13},\gamma_{14}\}=\{\alpha_{11}\otimes\alpha_{11},\alpha_{11}\otimes\alpha_{12},\alpha_{12}\otimes\alpha_{11},\alpha_{12}\otimes\alpha_{12}\},\nonumber\\
\gamma_{2}&=&\{\gamma_{21},\gamma_{22},\gamma_{23},\gamma_{24}\}=\{\alpha_{21}\otimes\alpha_{21},\alpha_{21}\otimes\alpha_{22},\alpha_{22}\otimes\alpha_{21},\alpha_{22}\otimes\alpha_{22}\},\nonumber\\
\gamma_{3}&=&\{\gamma_{31},\gamma_{32},\gamma_{33},\gamma_{34}\}=\{\alpha_{31}\otimes\alpha_{31},\alpha_{31}\otimes\alpha_{32},\alpha_{32}\otimes\alpha_{31},\alpha_{32}\otimes\alpha_{32}\}.\nonumber
\end{eqnarray}

Next, we will consider the relation of the coherence of quantum states in above AMUB.

A two-qubit Bell-diagonal states can be written as
\begin{eqnarray}\label{bs}
\rho_{B}=\frac{1}{4}(I\otimes I+\sum_{i=1}^3c_i\sigma_i\otimes\sigma_i),
\end{eqnarray}
where  $\{\sigma_i\}_{i=1}^3$ are the Pauli matrices, and $c_{1}, c_{2}, c_{3} \in[-1,1]$. The density matrix of  $\rho_{B}$ in base $\gamma_{1}=\{\gamma_{11},\gamma_{12},\gamma_{13},\gamma_{14}\}=\{|00\rangle$, $|01\rangle$, $|10\rangle$, $|11\rangle\}$ is:
\begin{eqnarray}\label{b1}
\rho_{B}= \frac{1}{4} \left(
\begin{array}{cccc}\label{bell}
1+c_3
& 0 & 0 & c_1-c_2 \\
0 & 1-c_3 & c_1+c_2 & 0 \\
0 & c_1 +c_2 & 1-c_3
& 0 \\
c_1-c_2 & 0 & 0 & 1+c_3
\end{array}
\right)
\end{eqnarray}
The $l_{1}$ norm of coherence of state $\rho_{B}$ in base $\gamma_{1}$ is
\begin{eqnarray}\label{bellc1}
C_{l_{1}}(\rho_{B})_{\gamma_{1}}=2(|\frac{1}{4}(c_{1}-c_{2})|+|\frac{1}{4}(c_{1}+c_{2})|)=\frac{1}{2}(|(c_{1}-c_{2})|+|(c_{1}+c_{2})|).
\end{eqnarray}

Let the density matrix of $\rho_{B}$ in base $\gamma_{2}=\{\gamma_{21},\gamma_{22},\gamma_{23},\gamma_{24}\}$ is
\begin{eqnarray}\label{b2}
\rho_{B}=\left(
\begin{array}{cccc}
d_{11} & d_{12} & d_{13} & d_{14} \\
d_{21} & d_{22} & d_{23} & d_{24} \\
d_{31} & d_{32} & d_{33} & d_{34}\\
d_{41} & d_{42} & d_{43} & d_{44}
\end{array}
\right),
\end{eqnarray}
and $\rho_{B}=d_{11}\gamma_{21}\gamma_{21}^{\dag}+d_{12}\gamma_{21}\gamma_{22}^{\dag}+d_{13}\gamma_{21}\gamma_{23}^{\dag}+d_{14}\gamma_{21}\gamma_{24}^{\dag}
+d_{21}\gamma_{22}\gamma_{21}^{\dag}+d_{22}\gamma_{22}\gamma_{22}^{\dag}+d_{23}\gamma_{22}\gamma_{23}^{\dag}+d_{24}\gamma_{22}\gamma_{24}^{\dag}
+d_{31}\gamma_{23}\gamma_{21}^{\dag}+d_{32}\gamma_{23}\gamma_{22}^{\dag}+d_{33}\gamma_{23}\gamma_{23}^{\dag}+d_{34}\gamma_{23}\gamma_{24}^{\dag}
+d_{41}\gamma_{24}\gamma_{21}^{\dag}+d_{42}\gamma_{24}\gamma_{22}^{\dag}+d_{43}\gamma_{24}\gamma_{23}^{\dag}+d_{44}\gamma_{24}\gamma_{24}^{\dag}.$
As $\rho_{B}$ in Eq. (\ref{b1}) and Eq. (\ref{b2}) is the same, using the method of undeterminated coefficients, we obtain
\begin{equation}
\begin{cases}
d_{11}+d_{12}+d_{13}+d_{14}+d_{21}+d_{22}+d_{23}+d_{24}+d_{31}+d_{32}+d_{33}+d_{34}+d_{41}+d_{42}+d_{43}+d_{44}=1+c_{3}\\
d_{11}-d_{12}+d_{13}-d_{14}+d_{21}-d_{22}+d_{23}-d_{24}+d_{31}-d_{32}+d_{33}-d_{34}+d_{41}-d_{42}+d_{43}-d_{44}=0\\
d_{11}+d_{12}-d_{13}-d_{14}+d_{21}+d_{22}-d_{23}-d_{24}+d_{31}+d_{32}-d_{33}-d_{34}+d_{41}+d_{42}-d_{43}-d_{44}=0\\
d_{11}-d_{12}-d_{13}+d_{14}+d_{21}-d_{22}-d_{23}+d_{24}+d_{31}-d_{32}-d_{33}+d_{34}+d_{41}-d_{42}-d_{43}+d_{44}=c_{1}-c_{2}\\
d_{11}+d_{12}+d_{13}+d_{14}-d_{21}-d_{22}-d_{23}-d_{24}+d_{31}+d_{32}+d_{33}+d_{34}-d_{41}-d_{42}-d_{43}-d_{44}=0\\
d_{11}-d_{12}+d_{13}-d_{14}-d_{21}+d_{22}-d_{23}+d_{24}+d_{31}-d_{32}+d_{33}-d_{34}-d_{41}+d_{42}-d_{43}+d_{44}=1-c_{3}\\
d_{11}+d_{12}-d_{13}-d_{14}-d_{21}-d_{22}+d_{23}+d_{24}+d_{31}+d_{32}-d_{33}-d_{34}-d_{41}-d_{42}+d_{43}+d_{44}=c_{1}+c_{2}\\
d_{11}-d_{12}-d_{13}+d_{14}-d_{21}+d_{22}+d_{23}-d_{24}+d_{31}-d_{32}-d_{33}+d_{34}-d_{41}+d_{42}+d_{43}-d_{44}=0\\
d_{11}+d_{12}+d_{13}+d_{14}+d_{21}+d_{22}+d_{23}+d_{24}-d_{31}-d_{32}-d_{33}-d_{34}-d_{41}-d_{42}-d_{43}-d_{44}=0\\
d_{11}-d_{12}+d_{13}-d_{14}+d_{21}-d_{22}+d_{23}-d_{24}-d_{31}+d_{32}-d_{33}+d_{34}-d_{41}+d_{42}-d_{43}+d_{44}=c_{1}+c_{2}\\
d_{11}+d_{12}-d_{13}-d_{14}+d_{21}+d_{22}-d_{23}-d_{24}-d_{31}-d_{32}+d_{33}+d_{34}-d_{41}-d_{42}+d_{43}+d_{44}=1-c_{3}\\
d_{11}-d_{12}-d_{13}+d_{14}+d_{21}-d_{22}-d_{23}+d_{24}-d_{31}+d_{32}+d_{33}-d_{34}-d_{41}+d_{42}+d_{43}-d_{44}=0\\
d_{11}+d_{12}+d_{13}+d_{14}-d_{21}-d_{22}-d_{23}-d_{24}-d_{31}-d_{32}-d_{33}-d_{34}+d_{41}+d_{42}+d_{43}+d_{44}=c_{1}-c_{2}\\
d_{11}-d_{12}+d_{13}-d_{14}-d_{21}+d_{22}-d_{23}+d_{24}-d_{31}+d_{32}-d_{33}+d_{34}+d_{41}-d_{42}+d_{43}-d_{44}=0\\
d_{11}+d_{12}-d_{13}-d_{14}-d_{21}-d_{22}+d_{23}+d_{24}-d_{31}-d_{32}+d_{33}+d_{34}+d_{41}+d_{42}-d_{43}-d_{44}=0\\
d_{11}-d_{12}-d_{13}+d_{14}-d_{21}+d_{22}+d_{23}-d_{24}-d_{31}+d_{32}+d_{33}-d_{34}+d_{41}-d_{42}-d_{43}+d_{44}=1+c_{3}
\end{cases}.
\end{equation}
The solution of the equation is
\begin{equation}
\begin{cases}
d_{11}=\frac{1+c_{1}}{4}, d_{12}=0, d_{13}=0,d_{14}=\frac{c_{3}-c_{2}}{4},\\
d_{21}=0, d_{22}=\frac{1-c_{1}}{4}, d_{23}=\frac{c_{3}+c_{2}}{4},d_{24}=0,\\
d_{31}=0, d_{32}=\frac{c_{3}+c_{2}}{4}, d_{23}=\frac{1-c_{1}}{4},d_{24}=0,\\
d_{41}=\frac{c_{3}-c_{2}}{4}, d_{42}=0, d_{43}=0,d_{44}=\frac{1+c_{1}}{4}.\\
\end{cases}
\end{equation}
So, the density matrix of  $\rho_{B}$ in base $\gamma_{2}=\{\gamma_{21},\gamma_{22},\gamma_{23},\gamma_{24}\}$ is
\begin{eqnarray}
\rho_{B}= \frac{1}{4} \left(
\begin{array}{cccc}
1+c_1 & 0 & 0 & c_3-c_2 \\
0 & 1-c_1 & c_3+c_2 & 0 \\
0 & c_3+c_2 & 1-c_1& 0 \\
c_3-c_2 & 0 & 0 & 1+c_1
\end{array}
\right).
\end{eqnarray}
The $l_{1}$ norm of coherence of state $\rho_{B}$ in base $\gamma_{2}$ is
\begin{eqnarray}\label{bellc2}
C_{l_{1}}(\rho_{B})_{\gamma_{2}}=2(|\frac{1}{4}(c_{3}-c_{2})|+|\frac{1}{4}(c_{3}+c_{2})|)=\frac{1}{2}(|(c_{3}-c_{2})|+|(c_{3}+c_{2})|).
\end{eqnarray}

Similarly, the density matrix of  $\rho_{B}$ in base $\gamma_{3}=\{\gamma_{31},\gamma_{32},\gamma_{33},\gamma_{34}\}$ is
\begin{eqnarray}
\rho_{B}= \frac{1}{4} \left(
\begin{array}{cccc}
1+c_2 & 0 & 0 & c_3-c_1 \\
0 & 1-c_2 & c_3+c_1 & 0 \\
0 & c_3+c_1 & 1-c_2& 0 \\
c_3-c_1 & 0 & 0 & 1+c_2
\end{array}
\right).
\end{eqnarray}
The $l_{1}$ norm of coherence of state $\rho_{B}$ in base $\gamma_{3}$ is
\begin{eqnarray}\label{bellc3}
C_{l_{1}}(\rho_{B})_{\gamma_{3}}=2(|\frac{1}{4}(c_{3}-c_{1})|+|\frac{1}{4}(c_{3}+c_{1})|)=\frac{1}{2}(|(c_{3}-c_{1})|+|(c_{3}+c_{1})|).
\end{eqnarray}

\begin{figure}[h]
\begin{center}
\scalebox{1.0}{\includegraphics[width=8cm]{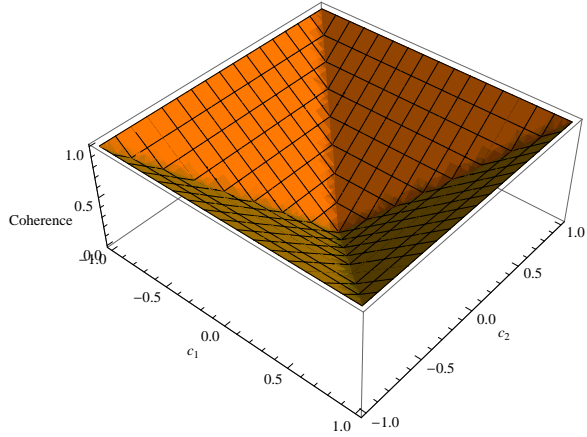}}
\caption{(Color online) The $l_{1}$ norm of coherence of Bell-diagonal states $\rho_{B}$ in base $\gamma_{1}$ as a function of $c_{1}$ and $c_{2}$.}
\end{center}
\end{figure}
In Fig. 1, the $l_{1}$ norm of coherence of Bell-diagonal states $\rho_{B}$ in base $\gamma_{1}$ as a function of $c_{1}$ and $c_{2}$ is depicted. When $c_{1}=c_{2}=0$, the coherence reach minimal value $0$. As $|c_{1}|$ and $|c_{2}|$ increase, the coherence increase. When $|c_{1}|=1$ or $|c_{2}|=1$, the coherence obtain maximum value. Similar situation also appear in the coherence in bases $\gamma_{2}$ and $\gamma_{3}$.

Next, we denote the sum of the $l_{1}$ norm of coherence of Bell-diagonal states $\rho_{B}$ in bases $\gamma_{1}$, $\gamma_{2}$, $\gamma_{3}$ by $C_{l_{1}}(\rho_{B})_{\gamma}$, i. e
\begin{eqnarray}
C_{l_{1}}(\rho_{B})_{\gamma}=C_{l_{1}}(\rho_{B})_{\gamma_{1}}+C_{l_{1}}(\rho_{B})_{\gamma_{2}}+C_{l_{1}}(\rho_{B})_{\gamma_{3}}.
\end{eqnarray}
In Fig. 2, we plot the surfaces \cite{lang} of the sum of the $l_{1}$ norm of coherence $C_{l_{1}}(\rho_{B})_{\gamma}$ of Bell-diagonal states $\rho_{B}$ in bases $\gamma_{1}$, $\gamma_{2}$, $\gamma_{3}$ in (a), (b), and (c). It show that the surface of the sum of the coherence is tetrahexahedron. As the sum increase, its volume expand, i. e. $|c_{1}|$, $|c_{2}|$, $|c_{3}|$ increase simultaneously.

\begin{figure}[h]
\begin{center}
\scalebox{1.0}{(a)}{\includegraphics[width=4.8cm]{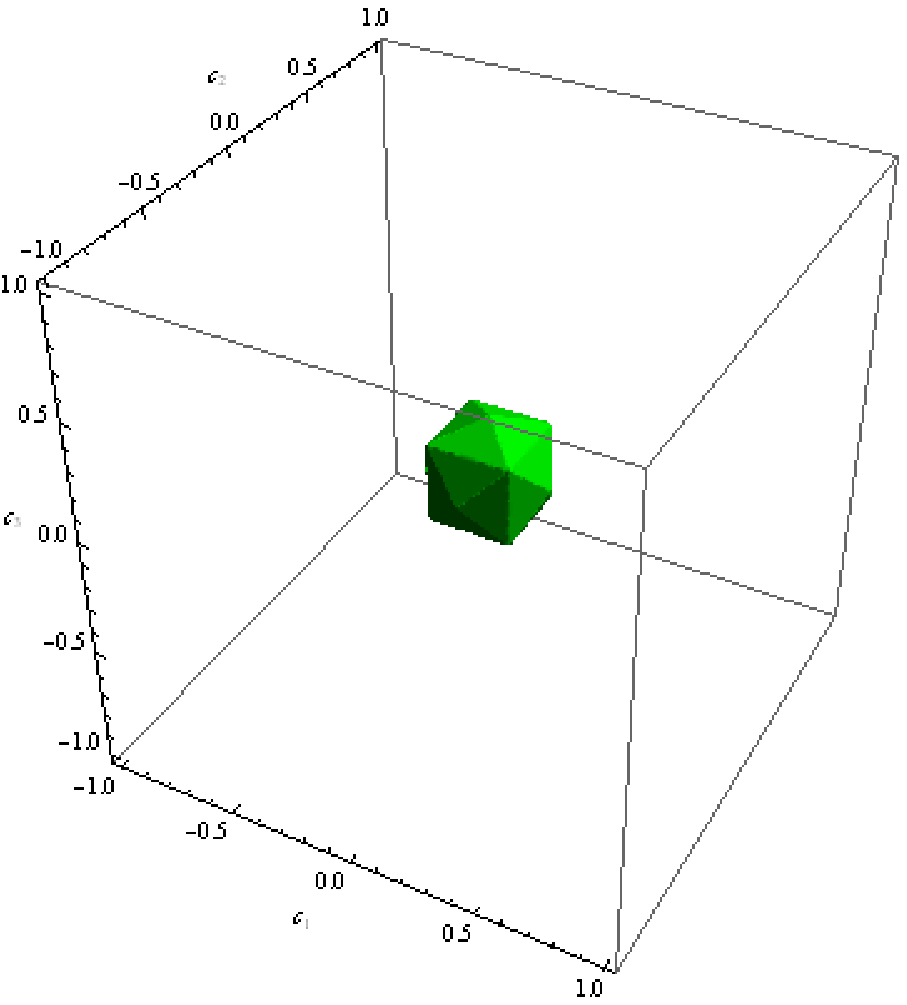}}
\scalebox{1.0}{(b)}{\includegraphics[width=4.8cm]{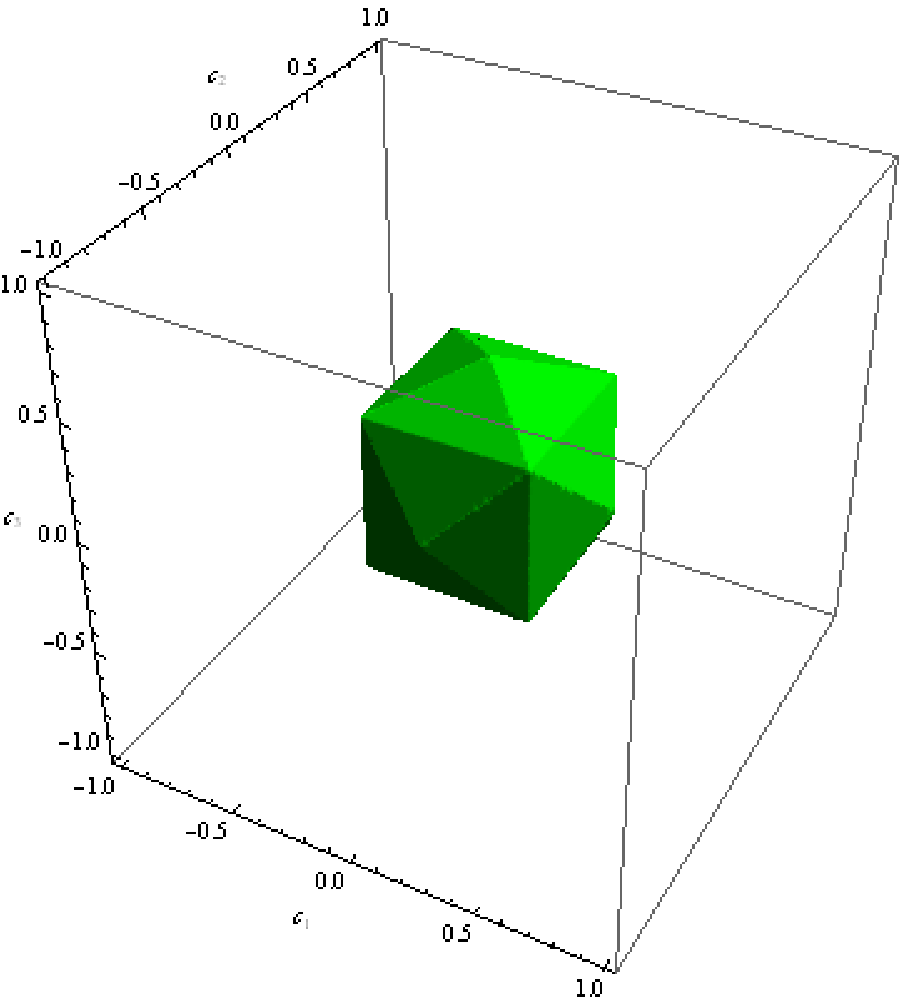}}
\scalebox{1.0}{(c)}{\includegraphics[width=4.8cm]{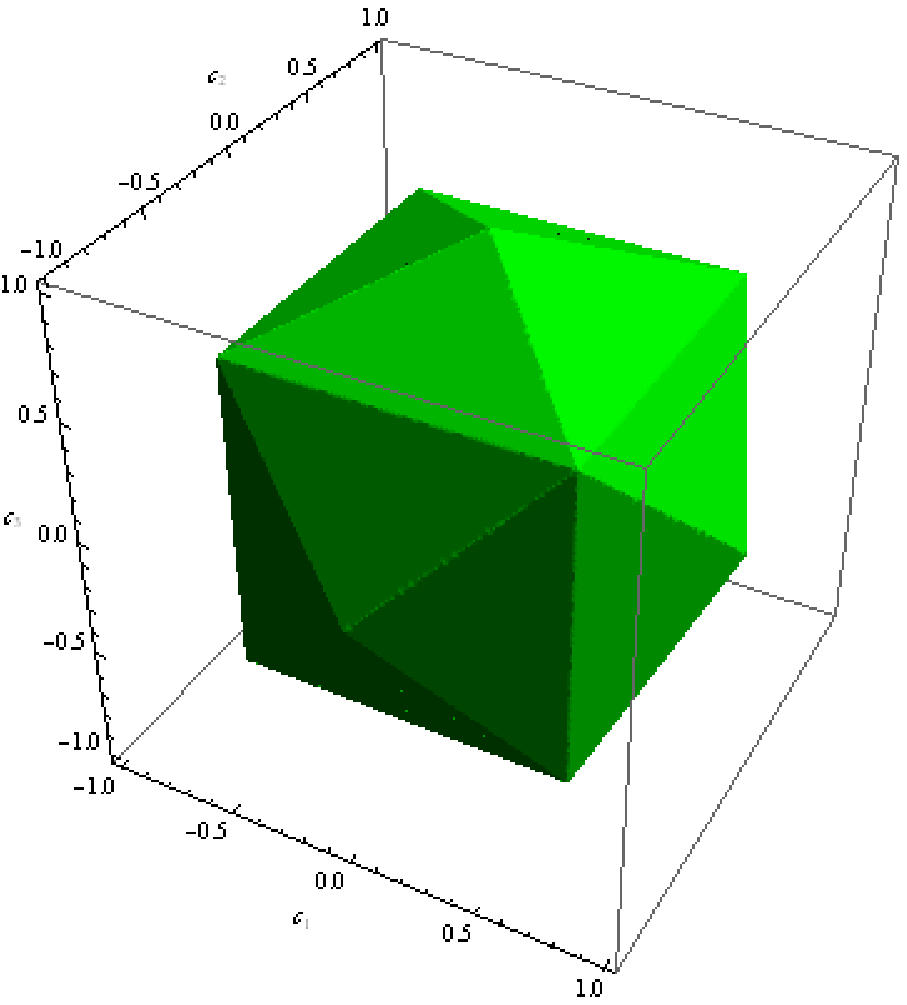}}
\caption{Surfaces of constant of the sum of the $l_{1}$ norm of coherence $C_{l_{1}}(\rho_{B})_{\gamma}$ for Bell-diagonal states $\rho_{B}$ in bases $\gamma_{1}$,
$\gamma_{2}$, $\gamma_{3}$ : (a) $C_{l_{1}}(\rho_{B})_{\gamma}=0.5$; (b) $C_{l_{1}}(\rho_{B})_{\gamma}=1$; $C_{l_{1}}(\rho_{B})_{\gamma}=2$.}
\end{center}
\end{figure}

In Eq. (\ref{bell}), let $c_{1}=c_{2}=c_{3}=\frac{4p}{3}-1$, where $0\leq p \leq 1$ ,Bell-diagonal states $\rho_{B}$ turn into Werner state
\begin{eqnarray}
\rho_{W}= \left(
\begin{array}{cccc}
\frac{p}{3} & 0 & 0 & 0 \\
0 & -\frac{p}{3}+\frac{1}{2} & \frac{2p}{3}-\frac{1}{2} & 0 \\
0 & \frac{2p}{3}-\frac{1}{2} & -\frac{p}{3}+\frac{1}{2} & 0 \\
0 & 0 & 0 & \frac{p}{3}
\end{array}
\right).
\end{eqnarray}
We denoted the $l_{1}$ norm of coherence of Werner states $\rho_{W}$ in bases $\gamma_{1}, \gamma_{2}, \gamma_{3}$ by $C_{l_{1}}(\rho_{W})_{\gamma_{1}}, C_{l_{1}}(\rho_{W})_{\gamma_{2}}, C_{l_{1}}(\rho_{W})_{\gamma_{3}}$ respectively. By Eqs. (\ref{bellc1}), (\ref{bellc2}), (\ref{bellc3}), we find that $C_{l_{1}}(\rho_{W})_{\gamma_{1}}=C_{l_{1}}(\rho_{W})_{\gamma_{2}}= C_{l_{1}}(\rho_{W})_{\gamma_{3}}=|\frac{4p}{3}-1|$.

In Eq. (\ref{bell}), let $c_{1}=\frac{4F-1}{3}, c_{2}=-\frac{4F-1}{3}, c_{3}=\frac{4F-1}{3}$, where $0\leq F \leq 1$ ,Bell-diagonal states $\rho_{B}$ turn into isotropic state
\begin{eqnarray}
\rho_{iso}= \left(
\begin{array}{cccc}
\frac{F}{3}+\frac{1}{6} & 0 & 0 & \frac{2F}{3}-\frac{1}{6} \\
0 & \frac{1}{3}-\frac{F}{3} & 0 & 0 \\
0 & 0 & \frac{1}{3}-\frac{F}{3} & 0 \\
 \frac{2F}{3}-\frac{1}{6}& 0 & 0 & \frac{F}{3}+\frac{1}{6}
\end{array}
\right).
\end{eqnarray}
We denoted the $l_{1}$ norm of coherence of isotropic states $\rho_{iso}$ in bases $\gamma_{1}, \gamma_{2}, \gamma_{3}$ by $C_{l_{1}}(\rho_{iso})_{\gamma_{1}}, C_{l_{1}}(\rho_{iso})_{\gamma_{2}}, C_{l_{1}}(\rho_{iso})_{\gamma_{3}}$ respectively. By Eqs. (\ref{bellc1}), (\ref{bellc2}), (\ref{bellc3}), we find that $C_{l_{1}}(\rho_{iso})_{\gamma_{1}}=C_{l_{1}}(\rho_{iso})_{\gamma_{2}}= C_{l_{1}}(\rho_{iso})_{\gamma_{3}}=|\frac{4F-1}{3}|$.

\section{\bf summary}\label{IIII}

In this work, we studied the $l_{1}$ norm of coherence of quantum states in mutually unbiased bases. We have found the sum of squared $l_{1}$ norm of coherence of the mixed state single qubit is less than two. We have obtained the $l_{1}$ norm of coherence of three classes of $X$ states in nontrivial mutually unbiased bases for $4$-dimensional Hilbert space is equal. We have proposed ``autotensor of mutually unbiased basis(AMUB)" by the tensor of mutually unbiased bases, and given the level surface\cite{lang} of constant the sum of the $l_{1}$ norm of coherence of Bell-diagonal states in AMUB. We have found the $l_{1}$ norm of coherence of Werner states and isotropic states in AMUB is equal respectively.

\bigskip

\end{document}